\documentclass[journal]{IEEEtran}
\ifCLASSINFOpdf
\else
\fi
\usepackage{graphicx}
\usepackage{caption}
\usepackage{amsmath}
\usepackage{bbm}
\usepackage{multirow}
\usepackage[numbers,sort&compress]{natbib}
\usepackage{subfigure}
\usepackage[hidelinks]{hyperref}
\usepackage[linesnumbered,ruled]{algorithm2e}
\usepackage{harpoon}
\usepackage{color}
\begin{document}
%
\title{Goal-Oriented UAV Communication Design and
Optimization for Target Tracking: A Machine Learning Approach}

\author{Wenchao Wu, \textit{Student Member, IEEE}
        Yanning Wu,
        Yuanqing Yang,
        and~Yansha Deng, \textit{Senior Member, IEEE} 
\thanks{Manuscript received 01 January 2024; revised 08 January 2024; revised 02 May; revised 10 July; accepted 07 August 2024. Date of publication 18 April 2024; This work was supported by the Engineering and Physical Sciences Research Council (EPSRC), U.K., under Grant EP/W004348/1.This work is also a contribution by Project REASON, a UK government-funded project under the FONRC sponsored by the DSIT. The associate editor coordinating the review of this letter and approving it for publication was Long D. Nguyen. (Corresponding author: Yansha Deng.)
\par The authors are with the Department of Engineering, King’s College London, WC2R 2LS London, U.K. (e-mail: wenchao.wu@kcl.ac.uk; yanning.wu@kcl.ac.uk; yuanqing.yang@kcl.ac.uk; yansha.deng@kcl.ac.uk).
}
}
        


%


\maketitle

\begin{abstract}
To accomplish various tasks, safe and smooth control of unmanned aerial vehicles (UAVs) needs to be guaranteed, which cannot be met by existing ultra-reliable low latency communications (URLLC). This has attracted the attention of the communication field, where most existing work mainly focused on optimizing communication performance (i.e., delay) and ignored the performance of the task (i.e., tracking accuracy). To explore the effectiveness of communication in completing a task, in this letter, we propose a goal-oriented communication framework adopting a deep reinforcement learning (DRL) algorithm with a proactive repetition scheme (DeepP) to optimize C\&C data selection and the maximum number of repetitions in a real-time target tracking task, where a base station (BS) controls a UAV to track a mobile target. The effectiveness of our proposed approach is validated by comparing it with the traditional proportional integral derivative (PID) algorithm.
\end{abstract}


%
\IEEEpeerreviewmaketitle

\vspace{-0.1in}
\section{Introduction}

\IEEEPARstart{D}{ue} to the high mobility, low cost, and line-of-sight communication, unmanned aerial vehicles (UAVs) have been widely used to accomplish various tasks, including image classification, parcel delivery, target detection, IoT management, and base station (BS) substitution \cite{UAV1, UAV2, UAV3, UAV4, UAV5, UAV6}. In order to successfully complete the task, the smooth and safe control of the UAV is important, which demands stringent quality of service (QoS) requirements (i.e., high reliability) for the downlink control and command (C\&C) data transmission. However, the ultra-reliable low-latency communications (URLLC) provided by the existing fifth-generation (5G) mobile communication network cannot meet such requirements. 
\par This challenge has raised increasing research interest from the communications field, where most existing research mainly focused on enhancing communication performance, such as latency, reliability, and data rate \cite{latency, reliability, data_rate}. In \cite{latency}, the optimal response delay was theoretically derived with close-form solutions for a swarm of three-dimensional distributed UAVs. In \cite{reliability}, the closed-form analytical expressions of the average packet error probability and effective throughput in the control link of UAV communications were formulated. To guarantee the data rate requirement and motion control performance of UAV, a data rate triggered sensing-control pattern activation in cellular-connected UAV networks was designed and its closed-form expression was obtained \cite{data_rate}. However, in a practical robotic task, the goal-oriented performance metric (i.e., tracking accuracy) is more important than communication metric. And there is a lack of research focused on improving it.

\par To fill this gap, goal-oriented communication (a new communication paradigm focuses on how the transmitted bits affect the goal) has been proposed in \cite{TO1} to design the communication system with a focus on the effectiveness of communication (the impact of the communication on the goal) in accomplishing a specific task. One promising approach to optimizing robotic task performance (i.e., classification accuracy \cite{Metrics}) is deep reinforcement learning (DRL). To generate optimal C\&C data in the UAV waypoint transmission task, DRL was applied at the BS to minimize the distance between the UAV's actual positions and the UAV's target positions at the end of each transmission time interval (TTI) \cite{TO2}. To reduce the redundant downlink C\&C data transmissions for the UAV, DRL was utilized at the BS with a focus on maximizing the task-oriented semantic-aware information \cite{TO3}. Though this framework has recently been proposed, there is only a little research using it to optimize goal-oriented performance metrics in a practical robotic control task.

\par Motivated by this, in this letter, we first propose a goal-oriented communication framework for the UAV downlink C\&C data transmission in a real-time target tracking task, where a BS controls a UAV to track a mobile target in real time. With the goal of maximizing the probability of successful tracking for this task, we propose a DRL algorithm along with a proactive repetition scheme (DeepP) to optimize the generation of the C\&C data and maximum repetition number. Compared with the traditional proportional integral derivative (PID) algorithm \cite{PID}, our proposed DeepP algorithm can increase the probability of successful tracking by 5.4 times.

\par The rest of this paper is organized as follows. Section II presents the system model and problem formulation. Section III introduces the DeepP algorithm. Section IV provides the simulation results. Finally, Section V concludes the paper.

\section{System Model and Problem Formulation}
In this section, a real-time target tracking task is introduced, where a BS sends C\&C data to control a UAV to track a mobile target in real time. Subsequently, we model the communication environment and introduce the proactive repetition scheme to transmit the downlink C\&C data. According to it, the problem formulation is presented.
\vspace{-0.1in}
\subsection{Real-time Target Tracking Task}
\begin{figure}[htbp]
    \vspace{-10pt}
    \centering
    \includegraphics[width=0.6\linewidth]{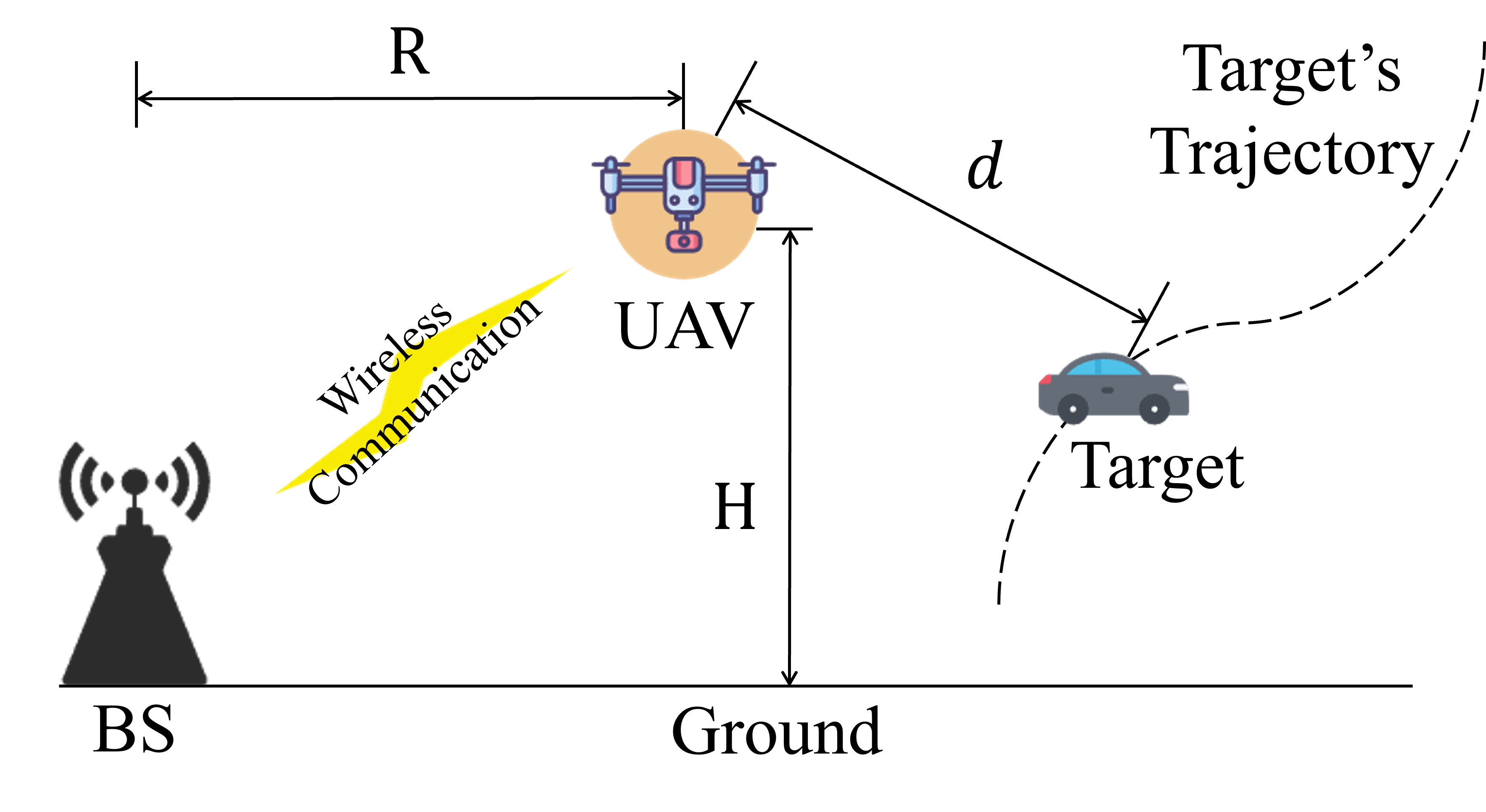}
    \caption{System model}
    \label{System model}
\end{figure}

Without loss of generality, we adopt a fundamental goal-oriented communication design in the real-time target tracking task shown in Fig. \ref{System model}, where a BS controls a UAV to track a mobile target (i.e., vehicle) moving alongside a random trajectory in real-time. This design can extend to multiple BS, UAV and mobile target scenarios. We assume the mobile target can send its real-time positions to the BS correctly all the time without error and the UAV's onboard sensor captures its real-time positions, forwarding them to the BS. Upon receiving these positions, the BS generates C\&C data to manipulate the UAV to track the target, which is achieved by managing the distance $d$ between the UAV and the target. If $d$ is not larger than a distance threshold $d^\mathrm{th}$ ($d\leq d^\mathrm{th}$), the UAV successfully tracks the target. Otherwise, the UAV fails to track the target. 
\vspace{-0.3in}
\subsection{Channel Model}
As shown in Fig. \ref{System model}, the BS is assumed to be in a fixed position and the UAV is envisaged to navigate a circular horizontal disk defined by radius $R$ and height $H$, where the C\&C data is transmitted via downlink transmission from the BS to the UAV.
Incorporating both potential line-of-sight (LoS) and non-line-of-sight (NLoS) scenarios, we employ free-space path loss and Rayleigh fading to formulate the path loss from the BS to the UAV as
\begin{equation}
    h^\mathrm{DL}=
    \left\{
    \begin{aligned}
        &{(\frac{4 \pi d^\mathrm{UB} f^\mathrm{DL}}{c})}^{\alpha} \eta_{\mathrm{LoS}}\beta, \quad P_{\mathrm{LoS}}\\
        &{(\frac{4 \pi d^\mathrm{UB} f^\mathrm{DL}}{c})}^{\alpha} \eta_{\mathrm{NLoS}}\beta, \ P_{\mathrm{NLoS}},
    \end{aligned} 
    \right
    .
\end{equation}
where $d^\mathrm{UB}$ represents the UAV-BS distance, $f^\mathrm{D}$ is the downlink transmission frequency, $c$ denotes the light's speed, $\alpha$ is path loss exponent, and $\beta \sim \mathcal{CN}(0,1)$ represents the Rayleigh small-scale fading. The $\eta_{\mathrm{LoS}}$ and $\eta_{\mathrm{NLoS}}$ are the path loss coefficients in LoS and NLoS scenarios, respectively. By calculating the angle of the UAV as $\theta^\mathrm{U}=\frac{180}{\pi}\arcsin\frac{\mathrm{H}}{d^\mathrm{UB}}$, we formulate the probability of LoS case $P_{\mathrm{LoS}}$ as
\begin{equation}
    P_{\mathrm{LoS}}=\frac{1}{1+C_1e^{-C_1(\theta-C_2)}},
\end{equation}
where $C_1$ and $C_2$ are positive constants corresponding to the environment. Then, we can derive the downlink channel as
\begin{equation}
    h^\mathrm{DC}= (P_{\mathrm{LoS}}\eta_{\mathrm{LoS}}+P_{\mathrm{NLoS}}\eta_{\mathrm{NLoS}})
        {(\frac{4 \pi d^\mathrm{UB} f^\mathrm{DL}}{c})}^{\alpha} \beta.
\end{equation}
According to it, the signal to noise ratio (SNR) is derived by
\begin{equation}
    \mathrm{SNR}= \frac{\mathrm{P}h^\mathrm{DC}}{\sigma^2},
\end{equation}
where the BS transmits the C\&C data in power $\mathrm{P}$ and the Additive White Gaussian Noise (AWGN) power is $\sigma^2$. By using Eq. (4), the delay of the C\&C data with size $N^\mathrm{CC}$ and bandwidth $B^\mathrm{CC}$ is formulated as
\begin{equation}
    t^\mathrm{Tr}=\frac{\mathrm{N^{CC}}}{B^\mathrm{CC}\log(\mathrm{SNR+1})}.
\end{equation}
At the UAV, the C\&C data is able to be successfully decoded when its SNR exceeds the threshold $\gamma^\mathrm{th}$, which is decided by the parameter $\delta^\mathrm{CC}$ defined as
\begin{equation}
    \delta^\mathrm{CC}=\begin{cases}
        0,\quad \mathrm{SNR} \leq \gamma^\mathrm{th}\\
        1,\quad \mathrm{SNR}>\gamma^\mathrm{th}.
    \end{cases}
\end{equation}
If $\delta^\mathrm{CC}$ is 1, the C\&C data is successfully decoded with the assumption that it can be fully recovered.
\vspace{-0.2in}
\subsection{Proactive repetition scheme}
At the start of $n^\mathrm{th}$ TTI $t_{n-1}$, where $n$ ranges in $\{1,2,..,N\}$ with $N \in \mathbbm{N}$ is the index of the final TTI, the C\&C data $\boldsymbol{m}_n$ is generated and transmitted from the BS to the UAV. Meanwhile, at the end of $n^\mathrm{th}$ TTI $t_n$, the UAV and target send their position $\boldsymbol{p}_{n}^\mathrm{U}=(x_n^\mathrm{U},y_n^\mathrm{U},z_n^\mathrm{U})$ and $\boldsymbol{p}_n^\mathrm{TG}=(x_n^\mathrm{TG},y_n^\mathrm{TG},z_n^\mathrm{TG})$ to the BS, respectively. The length of one TTI is $T$, and the C\&C data $\boldsymbol{m}_n$ is represented as 
\begin{equation}
    \boldsymbol{m}_n= (\boldsymbol{v}_n,\tau_n^\mathrm{E}),
\end{equation}
where $\boldsymbol{v}_n=(v_n^x,v_n^y,v_n^z)$ is the planned UAV's velocity vector for $n^\mathrm{th}$ TTI consisting of the velocities on the x, y, and z axis, respectively. After receiving $\boldsymbol{m}_n$, the UAV is assumed to execute this command for a fixed $\tau_n^\mathrm{E}$ period, with $\tau_n^\mathrm{E}=T$.
\par To achieve reliable C\&C data transmission, we introduce the proactive repetition scheme \cite{proactive_repetition}, where the BS transmits the same C\&C data for a maximum number of $K^\mathrm{max}$ repetitions and the time duration between two adjacent repetitions is $T^\mathrm{rep}$. Concretely, for the C\&C data $\boldsymbol{m}_n$, its $j^\mathrm{th}$ repetition ($j \in \{1,..,K^\mathrm{max}\}$) is transmitted at $t_{n,j}=t_{n-1}+(j-1)T^\mathrm{rep}$. For each repetition, the UAV will feedback the corresponding acknowledgement (ACK) or negative acknowledgement (NACK) to the BS, defined as
\begin{equation}
    \delta^\mathrm{ACK}_{n,j}=\begin{cases}
        0,\ \delta^\mathrm{CC}_{n,j}=0\\
        1,\ \delta^\mathrm{CC}_{n,j}=1,
    \end{cases}
\end{equation}
where $\delta^\mathrm{CC}_{n,j}$ is the detection state (i.e., success or failure) of the $j^\mathrm{th}$ repetition for $\boldsymbol{m}_n$. If $\delta^\mathrm{ACK}_{n,j}=1$, the UAV transmits an ACK back. Otherwise, the UAV sends a NACK. If the BS receives ACK, it will terminate repetitions earlier. With an emphasis on the downlink communication design, the uplink transmission (i.e., the target's position) is assumed to be ideal without packet loss or delay. Once the UAV successfully decodes one repetition, it will ignore other repetitions.  

\subsection{Problem Formulation}
At the time $t$ in whole process $[0,TN]$, we denoted the UAV's position and the target's position as $\boldsymbol{p}_{\frac{t}{T}}^\mathrm{U}$ and $\boldsymbol{p}_{\frac{t}{T}}^\mathrm{TG}$, respectively. For each TTI, we uniformly spilt it into $L$ parts with the timestamp of the $l^\mathrm{th}$ ($l\in \{1,...,L\})$ part in $n^\mathrm{th}$ TTI as $[(n-1)T+\frac{T}{L}l]$. According to it, we can obtain the UAV's position $\boldsymbol{p}_{(n-1+\frac{l}{L})}^\mathrm{U}$ and the target's position $\boldsymbol{p}_{(n-1+\frac{l}{L})}^\mathrm{GT}$, respectively. Based on that, we formulate the distance $d_{n,l}$ between the UAV and the target at the timestamp $[(n-1)T+\frac{T}{L}l]$ as
\begin{equation}
    d_{n,l}=||\boldsymbol{p}_{(n-1+\frac{l}{L})}^\mathrm{U}-\boldsymbol{p}_{(n-1+\frac{l}{L})}^\mathrm{TG}||,
\end{equation}
where $||.||$ is the Frobenius norm. Then, we define a function $V(d_{n,l})$ to value the distance $d_{n,l}$. If $d_{n,l}$ is not larger than the threshold $d^\mathrm{th}$ (successful tracking), $V(d_{n,l})$ reaches its maximum value of 1. Otherwise (failed tracking), $V(d_{n,l})$ decreases as the distance $d_{n,l}$ increases and its range is from -1 to 0. According to it, the function $V(d_{n,l})$ is formulated as
\begin{equation}
    V(d_{n,l})=\frac{[-f(d^\mathrm{th}-d_{n,l})+1)](e^{d^\mathrm{th}-d_{n,l}}-2)}{2}+1,
\end{equation}
where the function $f(.)$ is expressed as
\begin{equation}
    f(x)=\begin{cases}
        1,&x \geq 0\\
        -1,&x<0.
    \end{cases}
\end{equation}
To achieve the tracking task effectively, we aim to maximize the long-term value function $V(d_{n,l})$. Based on that, we can formulate the problem as  
\begin{equation}
\begin{split}
     &\mathcal{P}1: \max \lim_{L\rightarrow \infty}\frac{1}{NL}\sum_{\substack{n=1}}^{\substack{N}} \sum_{\substack{l=1}}^{\substack{L}}V(d_{n,l})\\
     &\mathrm{s.t.}\  N,L \in \mathbbm{N}.
\end{split}
\end{equation}

\section{DRL-Based Velocity and Maximum Repetition Number Selection}
Due to the varying significance of C\&C data in accomplishing our task and the diverse channel conditions for transmission, it is feasible to set different maximum repetition numbers $K^\mathrm{max}_n$ for each C\&C data $\boldsymbol{m}_n$. As a result, to address the problem in Eq. (12), we define the downlink C\&C data transmission action $A_n=\{v_n^x,v_n^y,v_n^z,K_n^\mathrm{max}\}$ to optimize the C\&C data $\boldsymbol{m}_n=(v_n^x,v_n^y,v_n^z)$ and $K_n^\mathrm{max}$ selection at the beginning of $n^\mathrm{th}$ TTI $t_{n-1}$. At the BS, the action $A_n$ is chosen by accessing all prior historical observations $O^\mathrm{H}_{n'}$ from the previous TTIs $n'\in\{1,...,n-1\}$. The set $O^\mathrm{H}_{n'}$ includes the UAV's position $\boldsymbol{p}_{n'-1}^\mathrm{U}$ and the target's position $\boldsymbol{p}_{n'-1}^\mathrm{TG}$. By incorporating all histories of the measurement $O^\mathrm{H}_{n'}$ and the action $A_{n'}$, the observation in $n^\mathrm{th}$ TTI can be denoted as $O_n=\{A_1,O^\mathrm{H}_1,...,A_{n-1},O^\mathrm{H}_{n-1}\}$.
\par To select the optimal action $A_n$ at the start of $n^\mathrm{th}$ TTI, we aim to maximize the long-term average reward $R_n$ associated with the distance $d_n$ between the UAV and the target at the end of $n^\mathrm{th}$ TTI $t_n$, which is formulated as
\begin{equation}
    d_{n}=||\boldsymbol{p}_{n}^\mathrm{U}-\boldsymbol{p}_{n}^\mathrm{TG}||,
\end{equation}
Based on that, $R_n$ is formulated as
\begin{equation}
    R_n=V(d_{n}).
\end{equation}
The optimization depends on choosing the action parameter $A_n$ based on the observation history $O_n$ concerning the stochastic policy $\pi$, and it is derived as
\begin{equation}
     \mathcal{P}2:\quad \max_{\pi(A_n|O_n)} \sum_{\substack{k=n}}^{\substack{\infty}}\gamma^{k-n}\mathbbm{E}_{\pi}[R_k],
\end{equation}
where $\gamma \in (0,1]$ represents the discount factor, accounting for the weighting of future TTIs. This process is a Markov decision process because its state $S_n=\{x_{n-1}^\mathrm{TG}-x_{n-1}^\mathrm{U},y_{n-1}^\mathrm{TG}-y_{n-1}^\mathrm{U},z_{n-1}^\mathrm{TG}-z_{n-1}^\mathrm{U}\}$ is only associated with its previous state and past action. Since the channel situation cannot be obtained, this introduces a partially observable markov decision process (POMDP) problem, which is generally intractable.
\begin{algorithm}[h]
\DontPrintSemicolon
    \SetAlgoLined
    \caption{DQN}
    \KwIn{Action space $\mathcal{A}$, $N^\mathrm{I}$,  $N^{\boldsymbol{\theta}}$, training parameters.}
    Initialization: $\boldsymbol{\theta}$, $\boldsymbol{\theta}^*$, replay memory $M$.\\
    \For{$Iteration\leftarrow1$ to $N^\mathrm{I}$}
    {
        \For{$n\leftarrow1$ to $N$}
        {
            Obtain $\boldsymbol{p}_{n-1}^\mathrm{U}$ and $\boldsymbol{p}_{n-1}^\mathrm{TG}$.\\
            Generate probability $p_\epsilon$.\\
            \eIf {$p_{\epsilon}<\epsilon$}
            {
                $A_n$ is random chosen from $\mathcal{A}$.\\
            }
            {
                $A_n=\arg\max_{A}Q(S_n,A;\boldsymbol{\theta}_n)$.\\
            }
            \For{$j\leftarrow1$ to $K^\mathrm{max}$}
            {
                \If{there is ACK back}
                {
                    Break.\\
                }
                Update the positions of the UAV and target.\\
            }
            Obtain $\boldsymbol{p}_{n}^\mathrm{U}$ and $\boldsymbol{p}_{n}^\mathrm{TG}$, get $S_{n+1}$ and $R_n$.\\
            Store the transition $(S_n,A_n,R_n,S_{n+1})$ in $M$.\\
            Sample transitions randomly from $M$.\\
            Calculate $\nabla L(\boldsymbol{\theta}_n)$ and
            update $\boldsymbol{\theta}$.\\
        }
        Update $\boldsymbol{\theta^*=\theta}$ every $N^{\boldsymbol{\theta}}$ episodes.\\
    }
\end{algorithm}
\vspace{-10pt}
\par To address the problem in (15), we propose the DRL-based approach with its ability to select the optimal C\&C data and $K^\mathrm{max}$ to achieve the task by considering the dynamic communication environment. The training process involves multiple episodes with every episode containing $N$ TTIs. In each episode, the selected action $A_n$ and current state $S_n$ are fed into the Q-network with the parameter vector $\boldsymbol{\theta}_n$, where the predicted value is calculated by using the function $Q(S_n,A_n;\boldsymbol{\theta}_n)$. After that, the next state $S_{n+1}$ and the reward $R_n$ are forwarded to the network which has the same structure as Q-network, namely target Q-network with the parameter vector $\boldsymbol{\theta}_n^*$, to calculate the target value. Subsequently, these results are processed to calculate the gradient of the loss function which derived as
\begin{multline}
    \nabla L(\boldsymbol{\theta}_n)=\mathbbm{E}_{S_n,A_n,R_n,S_{n+1}}[(R_n+\gamma \max_{A}Q(S_{n+1},A;\boldsymbol{\theta}^*_n)\\
    -Q(S_{n},A_n;\boldsymbol{\theta}_n))\nabla_{\boldsymbol{\theta}}Q(S_n,A_n;\boldsymbol{\theta}_n)],
\end{multline}
where $\boldsymbol{\theta}_n$ is updated by
\begin{equation}
    \boldsymbol{\theta}_{n+1}=\boldsymbol{\theta}_n-\lambda_{\mathrm{RMS}} \nabla L(\boldsymbol{\theta}_n),
\end{equation}
where $\lambda_{\mathrm{RMS}}$ is the RMSprop learning rate. It is crucial to note that the DQN updates $\boldsymbol{\theta}^*$ by copying $\boldsymbol{\theta}$ every $N^{\boldsymbol{\theta}}$ episodes. To achieve the trade-off between exploration and exploitation, the $\epsilon -$greedy approach is employed with $\epsilon \in [0,1]$. In each TTI, a probability is randomly generated by the agent and compared with $\epsilon$. If the probability is less than $\epsilon$, the agent randomly chooses an action. Otherwise, the agent chooses the optimal action. The implementation of the DQN algorithm is shown in $\mathbf{Algorithm\ 1}$, which has the time complexity of $O(NN^\mathrm{I})$.

\vspace{-0.1in}
\section{Simulation Results}
In this section, we present the simulation results of our proposed DeepP algorithm and compare it with the PID algorithm. The location of the BS is $(0\mathrm{m},0\mathrm{m},0\mathrm{m})$. The initial positions of the UAV and the target are $\boldsymbol{p}_0^\mathrm{U}=(69\mathrm{m},70\mathrm{m},50\mathrm{m})$ and $\boldsymbol{p}_0^\mathrm{TG}=(70\mathrm{m},70\mathrm{m},50\mathrm{m})$. The values of $T$ and $N$ are 1 ms and 100. $f^\mathrm{DL}$, $\gamma^\mathrm{th}$, $\sigma^2$, and $\mathrm{P}$ are set as 5 GHz, 5.5 dB, -104 dBm, and 18 dBm, respectively. In the DQN algorithm, $\epsilon$ is 1, the batch size is 32, $\lambda_\mathrm{RMS}$ is $10^{-4}$ and $\gamma$ is 0.1. The size of one C\&C data $N_\mathrm{CC}$ is set as 100 bytes. The velocities $v^x_n$, $v^y_n$, and $v^z_n$ are selected form the sets $\{-2000,-1500,...,2000\}$, $\{-2000,-1500,...,2000\}$, and $\{0\}$, respectively. The maximum repetition number $K_n^\mathrm{max}$ is selected from $\{1,2,..,10\}$. For the PID algorithm, we only use the proportional term with the proportional gain of 0.5. We obtain the results through the average value of 1000 simulations for each algorithm.
\begin{figure}[htbp]
    \vspace{-10pt}
    \centering
    \includegraphics[width=0.6\linewidth]{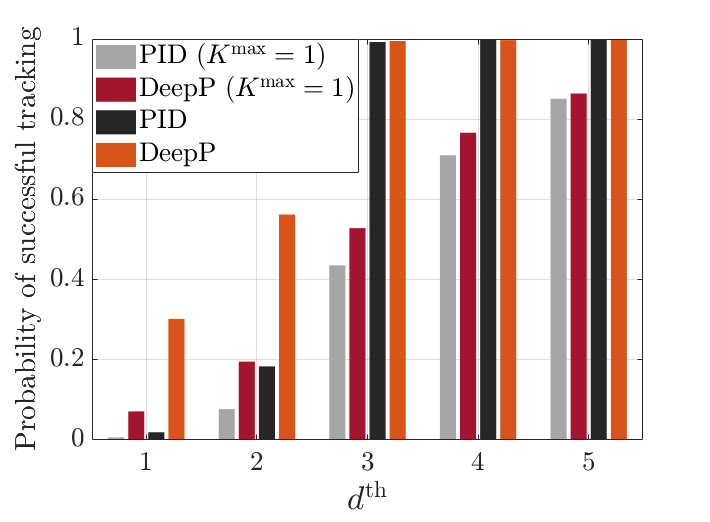}
    \vspace{-10pt}
    \caption{Performance comparison.}
    \label{Performance}
    \vspace{-10pt}
\end{figure}
\par Fig. \ref{Performance} plots the probability of successful tracking of the traditional PID algorithm when $K^\mathrm{max}=1$ and the $K^\mathrm{max}$ which approaches the performance limit, and also plots our proposed DeepP algorithm when $K^\mathrm{max}=1$ and $K^\mathrm{max}$ optimized by the DQN agent among different values of the threshold distance $d^\mathrm{th}$. We can obtain that as $d^\mathrm{th}$ increases, the probability of successful tracking of all algorithms in all cases increases. This is because the target tracking task can tolerate more packet loss when $d^\mathrm{th}$ is larger. It can also observed that our proposed DeepP algorithm outperforms the traditional PID algorithm in all cases, especially when there is a stringent $d^\mathrm{th}$ requirement. For example, when $d^\mathrm{th}=2$, our proposed DeepP algorithm can increase the probability of successful tracking by 1.54 times than the traditional PID algorithm when $K^\mathrm{max}=1$ and its probability of successful tracking is even higher than the traditional PID algorithm with the $K^\mathrm{max}$ approaching the performance limit. When DQN optimize $K^\mathrm{max}$, our proposed DeepP algorithm can increase the probability of successful tracking by 2.06 times. The results validate the effectiveness of our proposed DeepP algorithm.
\vspace{-0.1in}
\section{Conclusion}
In this letter, we designed a goal-oriented communication framework for the real-time target tracking task, where a BS controls a UAV to track the target in real-time. To increase the tracking success, we proposed a DeepP algorithm to select optimal C\&C data and $K^\mathrm{max}$ for the task. Our results shed light on that our proposed task-oriented communication framework can achieve a higher probability of successful tracking than the traditional PID algorithm.





%
\vspace{-0.1in}
\bibliographystyle{ieeetr}
\bibliography{mylib}

\end{document}